# Examinees' Rapid-Guessing Patterns in Computerized Adaptive Testing for Interim Assessment: From Hierarchical Clustering

Dandan Chen Kaptur[1]

Elizabeth Patton[2]

Logan Rome[3*]

[1] Pearson

[2] Meta Platforms Inc

[3] Curriculum Associates LLC

**Acknowledgments**

The authors thank the psychometrics and research teams at Curriculum Associates LLC for the support of this research initiative, and for their helpful feedback to improve the design of this study. In addition, the authors thank the editors and anonymous reviewers of this journal for their informative comments.

[*] Corresponding author; Email: lrome@cainc.com.

# Examinees' Rapid-Guessing Patterns in Computerized Adaptive Testing for Interim Assessment: From Hierarchical Clustering

## Abstract

Interim assessment is frequently administered via computerized adaptive testing (CAT), offering direct support to teaching and learning. This study attempted to fill a vital knowledge gap about the nuanced landscape of examinees' rapid-guessing patterns in CAT in the interim assessment context. We analyzed a sample of 146,519 examinees in Grades 1-8 who participated in a widely used CAT, using hierarchical clustering, a robust data science methodology for uncovering insights in data. We found that examinees' rapid-guessing patterns varied across item positions, content domains, chronological grades, examinee clusters, and examinees' overall rapid-guessing level on the test, suggesting a nuanced interplay between testing features and examinees' behavior. Our study contributes to the literature on rapid guessing in CATs for interim assessment, offering a comprehensive and nuanced pattern analysis and demonstrating the application of hierarchical clustering to process data analysis in testing.

## Introduction

Interim assessment may be considered low-stakes as it does not have direct consequences for examinees' academic or career opportunities. Due to the limited consequences associated with interim assessment scores, many examinees may not give their best effort when taking interim assessment. However, interim assessment may have long-term effects on examinees' learning and motivation, as it may reinforce effective learning, and thus may be perceived as high-stakes by educators and decisionmakers (Muskin, 2017). Non-effortful or disengaged responses reduce the number of informative items and contaminate the ability estimation (Wise

& Kuhfeld, 2019). These responses are more of a concern in interim assessment instead of high-stakes tests like SAT, because the low effort or disengagement is seen as an examinee's choice or responsibility to let go of the positive consequences in the high-stakes test (Wise & Kong, 2005). Identifying and addressing low-effort or disengaged responses is crucial for interim assessment because the presence of these responses, if ignored, can bias item parameters and lower score validity (Wise, 2017).

Examinees' low effort or disengagement can manifest in two forms: *rapid guessing*, characterized by unrealistically short response times, and *idle responding*, marked by prolonged duration when attempting test items (Gorgun & Bulut, 2023). In this article, we focus on examinees' rapid guessing instead of idle responding due to the relatively higher weight of rapid guessing to interim assessment. While both forms of disengagement are problematic, Gorgun and Bulut (2023) argued that idle responding is more concerning when speed and accuracy are measured together. This is especially true in reading and math assessments that require multiple skills at once (Samuels & Flor, 1997), or timed assessments that force examinees to balance speed and accuracy to avoid leaving items unanswered (e.g., Stickney et al., 2012). Interim assessment in many situations give students ample time, evaluating students based on their accuracy on the items they attempted, and hence rapid guessing would be a more relevant issue in interim assessment than idle responding (e.g., Gorgun & Bulut, 2023; Wise, 2006).

The evolving body of research on rapid guessing underscores the need for ongoing investigation and refinement of testing procedures to ensure robust assessment outcomes. Existing studies mainly examined rapid guessing when scrutinizing non-effortful or disengaged responses in interim assessment (e.g., Rose et al., 2010; Ulitzsch et al., 2020). Rapid-guessing behavior is the opposite of solution behavior, which is spending ample time seeking correct answers (Schnipke, 1995). The presence of rapid-guessing responses can undermine score validity (Wise, 2017), as examinees are scored based on the assumption that they have made their best effort when responding to each test item (Rios & Soland, 2021). Recent research shed light on the presence of rapid guessing across testing contexts and its implications for test score reliability and validity (e.g., Bolsinova & Tijmstra, 2018; Wang & Xu, 2015; Wise et al., 2021).

However, few studies have analyzed patterns in rapid guessing throughout item positions in computerized adaptive testing (CAT; Wainer, 2000). We hypothesize that rapid guessing may be influenced by features of the test design such as content domains, item positions, and between-domain activities (e.g., a game, a breathing exercise). We expect that test design and operation can benefit from understanding rapid-guessing patterns related to these factors.

CAT is a type of computer-based test that adapts to examinees' ability by selecting items based on an ability estimate calculated from their prior responses. It is widely used today to support classroom teaching and learning directly, such as *Star* (Renaissance Learning, 2020), *i-Ready Diagnostic* (Curriculum Associates, 2019), and *Map Growth* (NWEA, 2023). Rapid guessing is present in CAT-driven interim assessment (Finn, 2015). Lindner et al. (2019) explored how student characteristics were related to the onset of rapid guessing, but not the patterns of rapid guessing across item positions that may be influenced by features of the test.

This study is part of an effort to fill critical knowledge gaps about rapid-guessing patterns throughout CAT. Two research questions guided our exploration:

(1) What rapid-guessing patterns could be observed across item positions in interim assessment?

(2) How did rapid-guessing patterns correlate with test design features, such as content domains and breaks within interim assessment?

By identifying clusters of examinees based on their rapid guessing, we aim to provide insights for diagnostic purposes. These insights encompass pinpointing examinees who may require additional motivation or support, as well as selecting items that are more engaging for various types of test takers. Additionally, we seek to propose strategies to enhance test-taking engagement via test design.

To answer the research questions, we first quantified rapid guessing on each test item via an adapted normative threshold method, and then detected rapid-guessing patterns in the testing process via hierarchical clustering of rapid-guessed item data across item positions in a CAT. The data entails the accuracy and time of each item response from 146,519 Grades 1-8 students who took a popular CAT-based interim assessment in Math in Fall 2021. Our study centers on rapid-guessing behavioral changes across item positions in a CAT, offering a fresh and nuanced pattern analysis. The results demonstrated that data science techniques can effectively reveal patterns between examinees' rapid guessing behavior and the test design features of the CAT. These findings hold significant implications for improving CAT in interim assessment. While this study may not represent a definitive solution, it marks a promising initial step in this area of research. It offers valuable insights for test developers, administrators, and researchers on data science strategies for unveiling rapid-guessing patterns in CAT in interim assessment.

## Literature Review

### Effort, Engagement, and Rapid Guessing in Interim assessment

Score validity hinges upon examinees exerting maximal effort when addressing the items (American Educational Research Association et al., 2014). Effort, in this context, refers to individuals' investment in accomplishing a task (Inzlicht et al., 2018). Examinees' engagement during a test is reflected by their effort when attempting each test item. Solution behavior (Schnipke, 1995) is seen as *effortful* response behavior (Pastor et al., 2019), and rapid-guessing behavior is seen as *non-effortful* or *disengaged* response behavior (Kuhfeld & Soland, 2020).

One factor that may influence examinees' test-taking engagement is the time limit of the test. Timed tests are tests that have a fixed duration or a deadline for completion, while untimed tests are tests that allow examinees to take as much time as they need. Timed tests may induce speededness, which is the condition where examinees do not have enough time to answer all the items or have to rush through the test (Schnipke & Scrams, 1997). Speededness may lead to low-effort behavior, such as rapid guessing or skipping items. However, rapid guessing may also occur in untimed tests or interim assessment that provide sufficient time for examinees. In these cases, rapid guessing may reflect a lack of motivation, interest, or perceived value of the test (Wise & DeMars, 2005).

Specifically, rapid guessing, as one form of test-taking disengagement, is more common in low-stakes than high-stakes assessments (Finn, 2015). It attracts more attention in low-stakes interim assessments because rapid guessing in high-stakes tests is seen as an examinee's choice to forfeit the benefits of the test results, not much of a responsibility or concern for test developers or administrators (Wise & Kong, 2005). The presence of rapid guessing, when ignored, violates the local independence assumption and biases item- and person-parameter estimates (Wang & Xu, 2015). Including low-effort item responses can hamper score validity as it introduces construct-irrelevant variance (Wise, 2017); its impact can be large on individual test score estimates, although it is less of a problem for school-level aggregated scores and rankings, as shown in analyses with existing state-wide and international assessment data (Wise et al., 2020, 2021). Given the relatively high prevalence of rapid guessing in low-stakes tests, applying suitable methods to address the challenge it poses to score validity is of utmost importance for interim assessment.

**Approaches to Remedy Rapid Guessing in Interim Assessment**

Several remedies have been proposed to address rapid guessing in testing, but they vary in their effectiveness and feasibility. Many of them are reactive, meaning that they deal with disengagement after it occurs. For example, some methods suggest filtering out disengaged responses or examinees based on response time thresholds (e.g., Wang & Xu, 2015; Wise & DeMars, 2005; Wise & Kong, 2005) or self-report measures (e.g., Eklöf, 2006; Wise & Cotten, 2009). These methods flag and remove disengaged responses from the data after testing, which reduces the number of test items for scoring. Some methods suggest using an intelligent CAT that can detect and respond to disengagement (Gorgun & Bulut, 2023; Wise & Kingsbury, 2016; Wise, 2020), and implementing effort monitoring with proctor notification to intervene and re-engage test takers (Wise et al., 2019).

In addition, many studies attempted to improve the item- and person-parameter estimation by accounting for rapid guessing in CAT. In this regard, Wang and Xu (2015) proposed a mixture hierarchical model that improved the accuracy of item- and person-parameter estimates when accounting for both rapid-guessing behavior and solution behavior. To factor in the presence of local dependency between item responses and response times, Meng et al. (2015) proposed a conditional joint model with a covariance structure reflecting this dependency, Bolsinova et al. (2017) proposed to incorporate the residual response time effects into the existing item response theory (IRT) models' parameters, Bolsinova and Tijmstra (2018) proposed to integrate cross-loadings between item response time and person parameters in a hierarchical model, and Molenaar et al. (2016) proposed a hidden Markov approach to model item responses and response times after controlling for within-subject changes of an examinee's speed and ability in the test-taking process.

A proactive approach is needed to prevent or reduce disengaged responses during the test administration, which requires understanding the factors that influence test-taking engagement. The aforementioned reactive remedies may not be sufficient to maintain optimal testing conditions, as they may result in losing valuable information, reducing the precision of the ability estimates, or disrupting the testing process. In an extreme scenario, if an examinee displays rapid guessing for most of the items, it may not be possible to estimate the examinee's ability levels. A few studies have explored how examinee- or test-related characteristics may trigger disengagement in low-stakes settings. Tonidandel et al. (2002) argued that, in non-adaptive tests, the difficulty of the items may cause examinees to disengage if they are too hard or too easy for their ability level. Wise and Smith (2011) argued that it is the examinees' intrinsic motivation that may determine whether they invest effort or not. According to the Expectancy-Value Theory from Wigfield and Eccles (2000), another factor is the expected value of the task, which may depend on the perceived attainability and personal consequences of the item. An extensive review from Finn (2015) explored factors that may influence motivation and related methods to improve motivation. Incorporating these factors into the test design and operation may improve the test-taking engagement and performance of the examinees, as well as the quality and efficiency of the CAT.

**Methods to Identify Rapid Guessing**

Response time data for each item in CAT have been widely used as proxies to classify examinees' responses as rapid guessing. Various approaches have been explored to find time thresholds for flagging rapid guessing using item response times. Schnipke and Scrams (1997) observed that response times typically form a skewed distribution with an early spike due to rapid guessing. They suggested that rapid-guessing and solution behaviors have distinct time distributions and used a two-state mixture model to set item thresholds. Wise and Kong (2005) linked time thresholds to item length, positing that longer items should have higher thresholds. Wise (2006) recommended visually identifying thresholds by locating the end of the early spike in response time distributions. Kong et al. (2007) found that these methods yielded comparable thresholds.

However, setting thresholds in CAT is more complex due to the vast and dynamic item pools. Due to the large and dynamic nature of CAT item pools, setting thresholds for each item can be time-consuming and resource-intensive. Therefore, early CAT research often resorted to a uniform three-second threshold for its convenience and applicability, and yet this may not suffice for tests that include reading-intensive items (Wise & Ma, 2012). Kong et al. (2007) compared four methods and noted that variable thresholds performed slightly better than a uniform threshold in terms of agreement and validity. Wise and Ma (2012) proposed the normative threshold method, which defines an item's time threshold as a percentage (i.e., 10%, 15%, or 20%) of its median response time, up to a maximum of 10 seconds. This method can be easily applied to large and changing item pools.

The normative threshold method is not the only way to account for rapid guessing in CAT. Guo et al. (2016) proposed a nonparametric method, named the cumulative proportion (CUMP) method, finding response-time thresholds based on both response time and response accuracy. Rios and Guo (2020) proposed the mixture log-normal (MLN) method, fitting a mixed log normal distribution to the data and locate the threshold at the lowest point between two modes in a bimodal response time distribution. Bulut et al. (2023) explored the efficacy of a data-driven approach based on CUMP and MLN, compared with the normative threshold method using 10%, 15%, 20% and 25% of the median response time. In addition, many model-based approaches quantify rapid guessing using ability and speed (Guo et al., 2020; Ulitzsch et al., 2020, 2021). These studies indicate that rapid guessing is a complex phenomenon that requires careful and nuanced analysis.

The choice of a method for identifying rapid guesses in test data depends on many factors. It requires careful consideration of the tradeoffs between bias and precision, as well as the consequences for the users of the test. The normative threshold method is a preferable choice for identifying rapid guesses in test data, as it does not rely on any statistical assumptions, unlike MLN and CUMP methods (Soland et al., 2021) or other approaches based on probabilistic models. The normative threshold method also balances the tradeoffs between flagging too many or too few responses as rapid guesses, which can affect the validity and precision of the scores, as well as the operational implications for the test users (Wise et al., 2019; Wise & Ma, 2012). Moreover, it is flexible and adaptable to different purposes and uses of the scores, whether the scoring is for individual or aggregate-level decisions, and whether the scoring accounts for rapid guessing or not (Wise & Kuhfeld, 2020). It aligns with the purpose of this study to examine rapid-guessing patterns throughout the test. Therefore, the normative threshold method is reasonable and robust for detecting rapid guessing in this study.

**Hierarchical Clustering**

Hierarchical clustering is a versatile and powerful data science technique that groups observations to uncover hidden patterns. It is particularly effective for analyzing data with complex and non-flat structures, such as the response patterns of students in a CAT. Unlike other clustering methods, hierarchical clustering does not require assumptions such as linearity and local independence, and it can analyze complex data structures. It was chosen for this study due to its robustness, flexibility, and ease of interpretation.

Hierarchical clustering offers a more flexible alternative to K-Means clustering (Hartigan & Wong, 1979). K-Means clustering divides the data into a specified number of clusters, whereas hierarchical clustering results in a hierarchical structure of clusters without requiring specifying the number of clusters beforehand (Lisboa et al., 2013; Teichgraeber & Brandt, 2018). In addition, K-Means clustering is sensitive to outliers, as centroids used in this approach can be dragged by outliers (Franklin, 2019). Also, K-Means clustering struggles with data where clusters have various sizes or densities, as it requires the generalization of these clusters(*K-Means Advantages and Disadvantages*, 2022), whereas hierarchical clustering can effectively handle such variability (James et al., 2013). Moreover, K-Means clustering assumes a simple partitioning based on proximity to centroids, which does not help capture complex data structures that have hierarchies, whereas hierarchical clustering does not have this problem (Adams, 2018).

Hierarchical clustering does not assume linearity or reduce information as the principal component analysis (PCA; Hotelling, 1933) does. PCA is a commonly used technique for dimensionality reduction, which serves the purpose of clustering. However, PCA in essence identifies latent constructs or clusters that are a linear combination of multiple variables. In this process, there is information loss in dimensionality reduction, and variables are assumed to be linearly related to each other (McCoach et al., 2013). In contrast, hierarchical clustering does not require the stringent assumption of linearity, and it reflects the structure of all the data points, allowing for nesting structures and capturing complexities in the data effectively (Adams, 2018).

Hierarchical clustering does not come with stringent assumptions such as local independence as the latent class analysis (LCA; Lanza & Rhoades, 2013). LCA identifies latent classes or clusters based on a probabilistic model, which assumes local independence among observed variables. This means that, within each latent class or cluster, the observed variables are considered independent of one another once the class or cluster membership is accounted for (Hagenaars & McCutcheon, 2002). Hierarchical clustering, on the other hand, does not require such an assumption. It builds hierarchical structures based on observed distances or similarities between individual data points and between smaller clusters, without assuming a statistical model for the data. This feature makes hierarchical clustering a more generalizable method for identifying patterns in data.

Compared to some other methods for pattern analysis, hierarchical clustering also has its unique strengths. For instance, while DBSCAN (Ester et al., 1996) and spectral clustering (Ng et al., 2001) are valuable methods, they are not the best match for interim assessment data that could be hierarchical. Also, while DBSCAN, a density-based approach, is robust to outliers and does not require specifying the number of clusters, it is most suitable for geospatial data clustering and may struggle with data of varying densities. In contrast, hierarchical clustering does not have this problem and can provide a nuanced view of data in a hierarchical structure (Tate, 2023). While spectral clustering uses a graph-based approach that is great for complex clusters, it is nondeterministic due to its random component (Liu & Han, 2014). This characteristic means that it can produce different results each time it analyzes the same data set. Hierarchical clustering, on the contrary, is deterministic, generating the same results each time with the same data (Manning et al., 2008), which is favorable for reproducibility in analysis.

Agglomerative hierarchical clustering (Johnson, 1967) is the most commonly seen hierarchical clustering method, because of its computing efficiency (James et al., 2013). Specifically, when clustering data points, it will first treat each data point (i.e., an examinee) as a cluster, then cluster two data points at a time when they are identified with the smallest pairwise distance until all the data points are eventually assigned to a cluster. Next, it will cluster two clusters at a time until all the smaller clusters are assigned to a larger cluster. A convenient algorithm, the Lance-Williams (LW) algorithm (Lance & Williams, 1967), can be used to update inter-cluster distances in iterative computations, expressed as

$$d_{(ij)h} = \alpha_i d_{ih} + \alpha_j d_{jh} + \beta d_{ij} + \gamma |d_{ih} - d_{jh}|,$$

where

$d_{ih}$ denotes the distance between cluster $i$ and cluster $h$,

$(ij)$ denotes a new cluster, which is an agglomerate of cluster $i$ and cluster $j$; and

$\alpha_i$, $\alpha_j$, $\beta$ and $\gamma$ are parameters to be specified, depending on the linkage method.

Hierarchical clustering can use different distance measures and linkage criteria to quantify the similarity between clusters. Decisions on what distance measure to utilize and what linkage method to record inter-cluster distances can greatly impact the results from hierarchical clustering. With the complete linkage method, for instance, the maximum of the inter-cluster distances is recorded (as opposed to the minimum/mean in the single/average linkage method). The inter-cluster distance is then expressed as $d_{(ij)h} = \max \{d_{ih} + d_{jh}\}$, which can be translated into $d_{(ij)h} = \frac{1}{2} d(i, h) + \frac{1}{2} d(j, h) + \frac{1}{2} |d(i, h) - d(j, h)|$ using the LW update formula (Johnson, 1967), where $\alpha_i = \frac{1}{2}$, $\alpha_j = \frac{1}{2}$, $\beta = 0$, and $\gamma = \frac{1}{2}$. Table 1 summarizes the coefficient values for different linkage methods (Tan et al., 2005):

**Table 1.** *Coefficient Values for Different Linkage Methods*

| Linkage Method | $\alpha_i$ | $\alpha_j$ | $\beta$ | $\gamma$ |
|---|---|---|---|---|
| Complete Linkage | $\dfrac{1}{2}$ | $\dfrac{1}{2}$ | 0 | $\dfrac{1}{2}$ |
| Single Linkage | $\dfrac{1}{2}$ | $\dfrac{1}{2}$ | 0 | $-\dfrac{1}{2}$ |
| Average Linkage | $\dfrac{n_i}{n_i + n_j}$ | $\dfrac{n_i}{n_i + n_j}$ | 0 | 0 |

Dendrograms in hierarchical clustering are tree-like diagrams that help visualize the data structure and determine the number of meaningful clusters (Murtagh & Contreras, 2011). Figure 1 shows an example of dendrograms we created given the hierarchical clustering results based on the data from one grade of examinees. The bottom of the figure has a large number of straight lines, which are "leaves" corresponding to individual examinees. Moving up in the dendrogram, the "leaves" converge at "branches," and lower level "branches" fuse into higher level "branches." When cutting this dendrogram horizontally at a certain point, the "branches" below this cut are interpreted as clusters. The higher the fusion occurs, the fewer the clusters we get, and the farther the "leaves" are to each other between clusters. One dendrogram can result in many numbers of clusters, and it is up to researchers to decide on an appropriate number of clusters given the research purpose (James et al., 2013). This approach allows for the identification of meaningful clusters within the hierarchical structure by visually assessing where to make the cut.

Hierarchical clustering, while a powerful analytical tool, is not without its challenges. It involves arbitrary decisions regarding linkage methods, struggles with missing data, may falter when dealing with mixed data types, and is less effective for analyzing exceedingly large datasets (James et al., 2013). However, when it comes to analyzing examinees' data in CAT, hierarchical clustering proves to be particularly advantageous. The test data always exhibits a natural hierarchical structure, as item responses are nested within examinees. Hierarchical clustering excels at capturing this intricate structure.

We wanted to ensure that our method aligns with the data's inherent hierarchy and to guarantee the reproducibility of our analytical outcomes—without the restrictive assumptions of linearity or local independence. Hierarchical clustering is an optimal option given these considerations for our analysis, and hence it was chosen for our analysis. To address the limitations of hierarchical clustering as described above, we explored the use of three linkage methods during our preliminary analysis, which informed our selection of the most suitable linkage method.

## Method

### Test Design

The design of the selected CAT was sequential, with items from one domain presented consecutively before transitioning to the next. The test began with 18 items from Domain 1, followed by 20 items from Domain 2, then 14 items from Domain 3, and concluded with 14 items from Domain 4. This structured approach ensured that all items within a single domain were completed before moving on to the next, allowing for a focused assessment of each content area. Between the completion of one domain and the start of the next domain, examinees were provided with a "break" in the form of a game or a breathing exercise, depending on the settings chosen by their school or district.

The CAT was powered by a sophisticated algorithm, drawing test items adaptively from a vast item pool exceeding 5,000 items, each calibrated using the Rasch model (Rasch, 1960). The items manifest in a wide array of formats, encompassing multiple-choice, constructed-response, text-highlight, drag-and-drop, coordinate-grid, and number-line questions. The adaptive nature of the CAT ensured a personalized testing experience, with each examinee encountering different items at each item position. The selection of the subsequent item was determined by the alignment between the interim ability—derived from the examinee's responses to preceding items—and the difficulty level of the potential next item. After each response, the examinee's ability was updated via maximum likelihood estimation (MLE; Fisher & Russell, 1922) method.

The test was delivered through a user-friendly computer interface, presenting one item at a time in a large, clear font, thus enhancing readability and minimizing distractions. This format allowed examinees to concentrate fully on each item. It also has the option for examinees to divide the test into several sittings if needed, thereby accommodating diverse testing preferences and schedules.

**Data**

The original data comprised the response time for each item, for an extensive group of 146,519 students from Grades 1-8. They participated in a CAT in Math, conducted as part of interim assessment in the Fall term of 2021. To uncover rapid-guessing patterns among examinees, we scrutinized their response time data. This work involved creating binary "rapid-guessing indicators" for each item to detect rapid guessing. Also, it involved classifying examinees based on these indicators into classifications of "high" or "moderate" rapid guessing, facilitating a nuanced understanding of rapid guessing for examinees that varied in their overall effort on the CAT.

The distribution of the examinees in the retained data varied by grade, with the smallest group being Grade 6, consisting of 11,179 students, and the largest being Grade 1, with a notable 36,665 students. To validate the consistency and reliability of our findings, we partitioned the dataset into two nearly equal, random subsets, designated as "Set 1" and "Set 2." We then cross-checked the results obtained from all the data against those from "Set 1" and "Set 2" to ensure the robustness of our analysis. Detailed in Table 2 are the grade-specific sample sizes that constitute our data, providing a clear breakdown of examinees included in this study.

**Table 2.** *Number of Examinees by Grade per Dataset*

| Grade | Rapid-Guessing Classification | Number of Examinees | | |
|---|---|---|---|---|
| | | All | Set 1 | Set 2 |
| 1 | Moderate | 28931 | 14469 | 14538 |
| | High | 7734 | 3846 | 3871 |
| 2 | Moderate | 22731 | 11344 | 11377 |
| | High | 7017 | 3506 | 3485 |
| 3 | Moderate | 12059 | 6067 | 5992 |
| | High | 3787 | 1866 | 1922 |
| 4 | Moderate | 10383 | 5182 | 5202 |
| | High | 3668 | 1845 | 1824 |
| 5 | Moderate | 8800 | 3739 | 3787 |
| | High | 3470 | 1843 | 1810 |
| 6 | Moderate | 7526 | 5182 | 5202 |
| | High | 3653 | 1845 | 1824 |
| 7 | Moderate | 8063 | 3739 | 3787 |
| | High | 4550 | 1843 | 1810 |
| 8 | Moderate | 8569 | 4354 | 4207 |
| | High | 5578 | 2704 | 2869 |

Specifically, drawing from millions of examinees across thousands of elementary and middle schools, the examinees were chosen based on four criteria: (1) having completed the test within the school environment, (2) having completed the test in one single sitting, (3) falling within the interquartile range of the total number of items flagged for rapid guessing (i.e., marked with 1 in the corresponding rapid guessing indicators); and (4) being classified as exhibiting "high rapid guessing" or "moderate rapid guessing." Criterion (1) ensures the exclusion of any non-school-related environmental factors, as the data were collected during the pandemic when many students were allowed to take the test at home and self-report where they took the test (i.e., at home vs. in school). Criterion (2) addresses potential confounding factors due to multiple test sittings (i.e., related to the test administration, as opposed to test design). Criterion (3) minimizes the impact of extreme values and ensures that our analysis focuses on the most representative and typical response behaviors in this CAT. Criterion (4) allows us to concentrate on the rapid-guessing patterns central to our study. Meeting these criteria is paramount in assembling a dataset that is both comprehensive and conducive to producing reliable, high-quality insights.

**Measuring Rapid Guessing**

To explore how test design features may influence rapid guessing, we focused on how examinees' rapid guessing varied across different item positions in CAT. This approach is novel since most (if not all) of the previous studies on rapid guessing examined rapid guessing at the item- or test-level. We adapted measures for quantifying rapid guessing for individual items from the existing literature to suit this study. Below are the details.

We relied on the normative threshold method (Wise & Ma, 2012), due to its merits over other methods we covered in Literature Review. However, we made some minor modifications to this method to suit the specific characteristics of the CAT data we examined, based on preliminary findings from internal research conducted by the testing company that developed this CAT product. Specifically, we computed the item response time threshold, $T$, for each item, by multiplying the median response time at each item position by 0.1. We then compared the examinee's response time, in seconds, for each item to this threshold and generated a "rapid-guessing indicator" for that item. This indicator was coded as 1 if the response time was below

*T*, indicating rapid guessing, and as 0 otherwise. We opted for a 10% threshold instead of 20% or another higher threshold as done by some other studies (e.g., Bulut et al., 2023), because we wanted to be conservative and minimize the Type-I errors (i.e., falsely flagging a response as a rapid guess). Higher thresholds, such as 15% or 20%, could increase the Type-I error rates.

The structure of the data is displayed in Table 3. For example, as depicted in Table 3, examinees 1, 3, and 146,518 demonstrated a tendency towards rapid guessing at the onset of the test, as evidenced by a concentration of 1s in the initial four indicators. Conversely, examinees 2 and 146,519 exhibited a propensity for rapid guessing toward the end of the test, demonstrated by a series of 1s in the final three to four indicators. This structured approach to indicator analysis was crucial for identifying and understanding the various rapid-guessing behaviors present within the examinee population.

**Table 3.** *Binary Data Indicating Rapid Guessing per Item Position and per Examinee*

| Examinee | I1 | I2 | I3 | I4 | … | I63 | I64 | I65 | I66 |
|----------|-----|-----|-----|-----|-----|-----|-----|-----|-----|
| 1 | 0 | 1 | 1 | 1 | … | 0 | 0 | 0 | 0 |
| 2 | 0 | 0 | 0 | 0 | … | 0 | 1 | 1 | 1 |
| 3 | 1 | 1 | 1 | 0 | … | 0 | 0 | 0 | 0 |
| … | … | … | … | … | … | … | … | … | … |
| 146518 | 1 | 0 | 1 | 1 | … | 0 | 0 | 0 | 0 |
| 146519 | 0 | 0 | 0 | 1 | … | 1 | 1 | 1 | 1 |

*Note.* I1-I66 denote a set of rapid guessing indicators for each item position, from the 1st to the 66[th] position. The value 1 indicates an examinee's item response is "rapid guessing," and the value 0 indicates it is not.

Also, we aimed to quantify rapid-guessing patterns for different groups of examinees, taking into account the individual differences in rapid-guessing rates discussed in the Literature Review (e.g., Kröhne et al., 2020). To do this, we used two criteria to classify examinees: their response time effort (RTE; Wise & Kong, 2005) and proportion correct. These criteria reflect the speed and accuracy of responses, which are important factors to consider when flagging rapid guessing (Deribo et al., 2023; Schnipke & Scrams, 1997; Wise & Ma, 2012). RTE is the proportion of rapid-guessing indicators coded with 1 out of the total number of items for an examinee. We classified an examinee with an RTE below 0.75 and a proportion correct below 0.43 as exhibiting "high rapid guessing." We categorized an examinee with an RTE below 0.90 and a proportion correct below 0.45, but not meeting the criteria for "high rapid guessing" as demonstrating "moderate rapid guessing." These categories were based on rigorous internal research by the testing company that developed the CAT, using historical performance data from several years. This ensured that the categories were accurate and relevant to the CAT context.

**Analysis**

We used agglomerative hierarchical clustering to identify rapid-guessing patterns due to its merits we covered in Literature Review. As we focused on examinee-centric rapid-guessing patterns in association with test design features, our unit of analysis was examinees' binary rapid-guessing indicators for all the 66 item positions; the objective was to put individual examinees into interpretable clusters and uncover discernable rapid-guessing patterns for each cluster. We performed the analysis by chronological grade (i.e., Grades 1-8), given our assumption that rapid-guessing patterns could vary across grades, and that grade-specific findings would be most beneficial to future test development and administration that often vary for each grade. The package "stats" and the software R 4.2.1 (R Core Team, 2021) were utilized for this analysis.

To capture the change of rapid guessing, we calculated the proportions of rapid-guessing occurrences (i.e., 1s of rapid-guessing indicators) at each item position per grade, per the overall rapid-guessing level on the test, and per cluster. For visual inspection, we plotted these proportions on the $y$-axis against item positions on the $x$-axis (i.e., the 1st, 2nd, up to the 66th item position), and linked them together into rapid-guessing trajectories. This proportion on the

*y*-axis was calculated by dividing the number of responses flagged for rapid guessing by the total number of responses. Figure 1 shows an example of the rapid-guessing trajectories for examinees in one grade. Item positions 1-18 were for items in Domain 1, positions 19-38 for Domain 2, positions 39-52 for Domain 3, and positions 53-66 for Domain 4.

In essence, this proportion reflects how the rapid-guessing rate at each item position varied across the test for each cluster of examinees. The trajectory plot made it possible to inspect the prevalence of rapid guessing throughout the test: A horizontal trajectory across positions would suggest that examinees' rapid guessing was not related to the item position, whereas any inclination in the trajectory would indicate a systematic relationship between low effort and the item position.

We made a few decisions at each step of our analysis, based on the following principles: parsimony for visualization, consistency for interpretation, and maximized difference between clusters. First, we did not standardize the binary rapid-guessing indicators to facilitate interpretation. Second, we determined the optimal number of clusters using dendrograms from the hierarchical clustering. In this step, for example, for both the two categories of examinees (i.e., "moderate rapid guessing" on the left side, "high rapid guessing" on the right side) in one grade, shown in Figure 2, it was possible to cut both their dendrograms horizontally at different positions, showing that the number of clusters could be either two or three. We decided to move forward with two clusters instead of three clusters,

**Figure 1.** *Rapid-Guessing Trajectories (One Grade)*

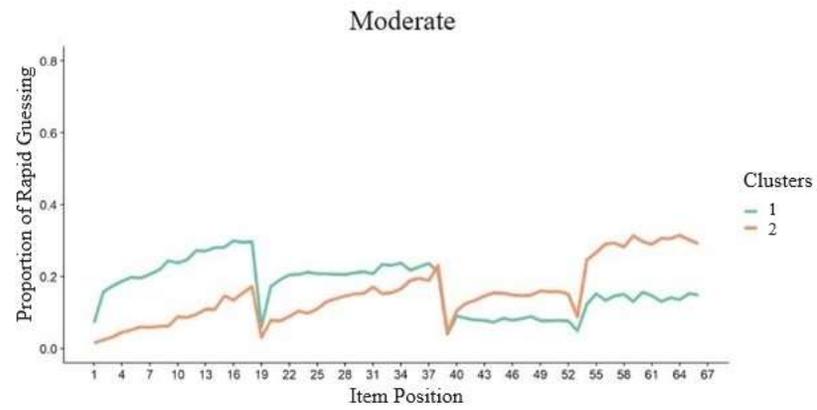
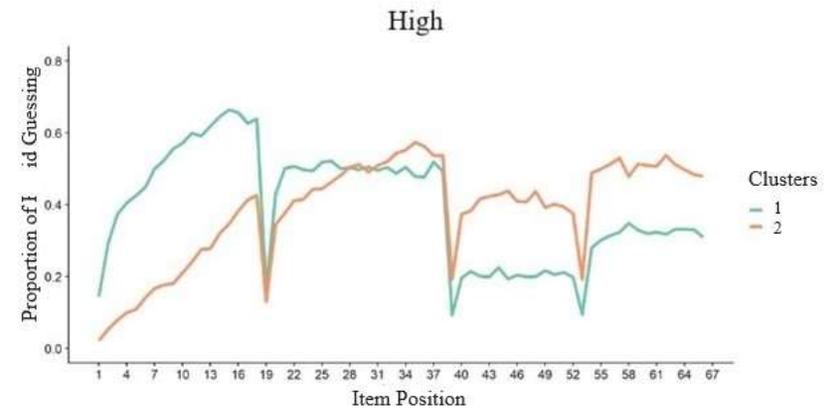

**Figure 2.** *Dendrograms to Determine the Number of Clusters (One Grade)*

**Figure 3.** *Silhouette Coefficient to Determine the Number of Clusters (One Grade)*

Moderate                                                    High

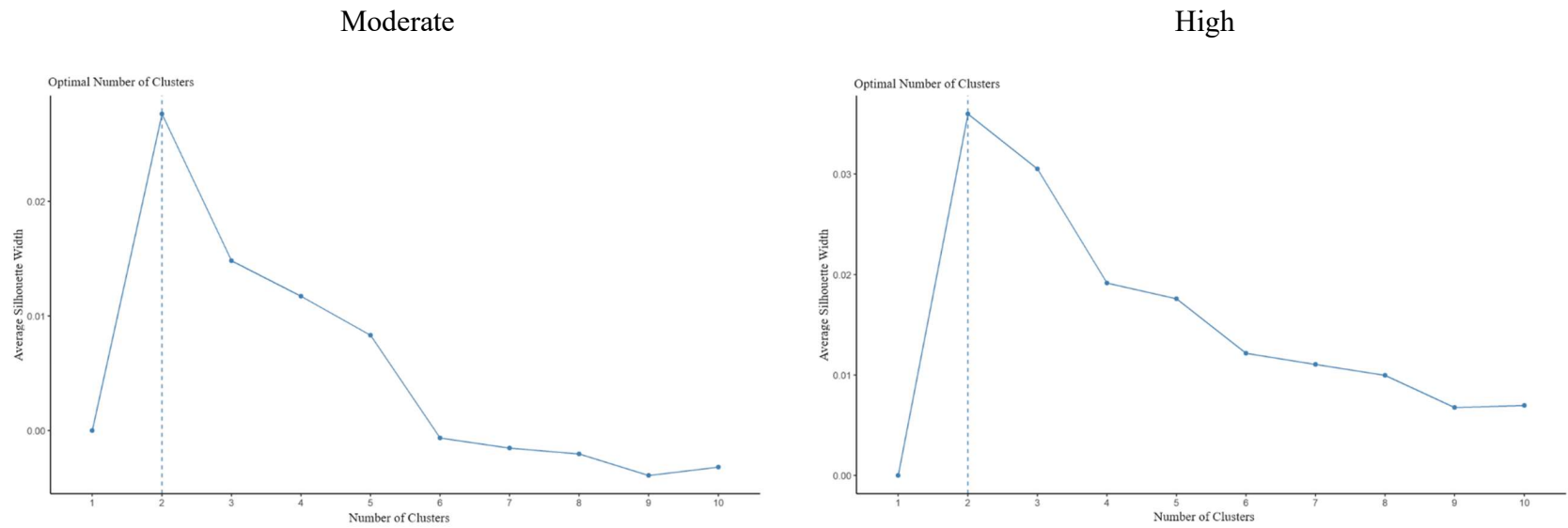

because we found that using two clusters would ensure relatively more parsimonious, consistent and interpretable rapid-guessing patterns across grades, and would maximize the distance between clusters, as two was associated with a higher fusion in dendrograms than three. Also, to cross-check this number of clusters, we used the silhouette coefficient to evaluate the number of clusters, shown in Figure 3, which also indicated that we should consider two clusters instead of three. Third, we ran hierarchical clustering using this number of clusters. This clustering eventually coded examinees into two clusters.

In this third step, we utilized Euclidean distance to quantify dissimilarity and the complete linkage method to quantify the inter-cluster distance. We chose Euclidean distance because this measure is suitable for binary data (Lourenço et al., 2004) and can capture the magnitude of changes (James et al., 2013), which met our needs. Euclidean distance is the square root of the sum of the squared differences between data points, expressed as $d_{pq} = \sqrt{\sum_{i=1}^{n}(q_i - p_i)^2}$ where $p$ and $q$ are two points, and $n$ denotes the dimension of the sample space. The choice of the complete linkage method was based on our early exploration with each of three popular linkage methods (i.e., average, complete, single linkage methods), which unveiled that the patterns from this linkage method were the most parsimonious and interpretable, as shown in Figure 4.

We performed hierarchical clustering and created rapid-guessing trajectories for the two categories of examinees (i.e., "high" or "moderate" rapid guessing) separately for two reasons. First, we hypothesized that the rapid-guessing patterns would vary across the levels, and we wanted to show a nuanced understanding of rapid guessing for examinees with different effort levels on the CAT. Second, we found a lot of noise in the rapid-guessing patterns in the aggregated data, as seen by many zigzags on the left side of Figure 5, and little difference in

**Figure 4.** *Patterns from Three Linkage Methods (One Grade)*

- Average Linkage

- Complete Linkage

- Single Linkage

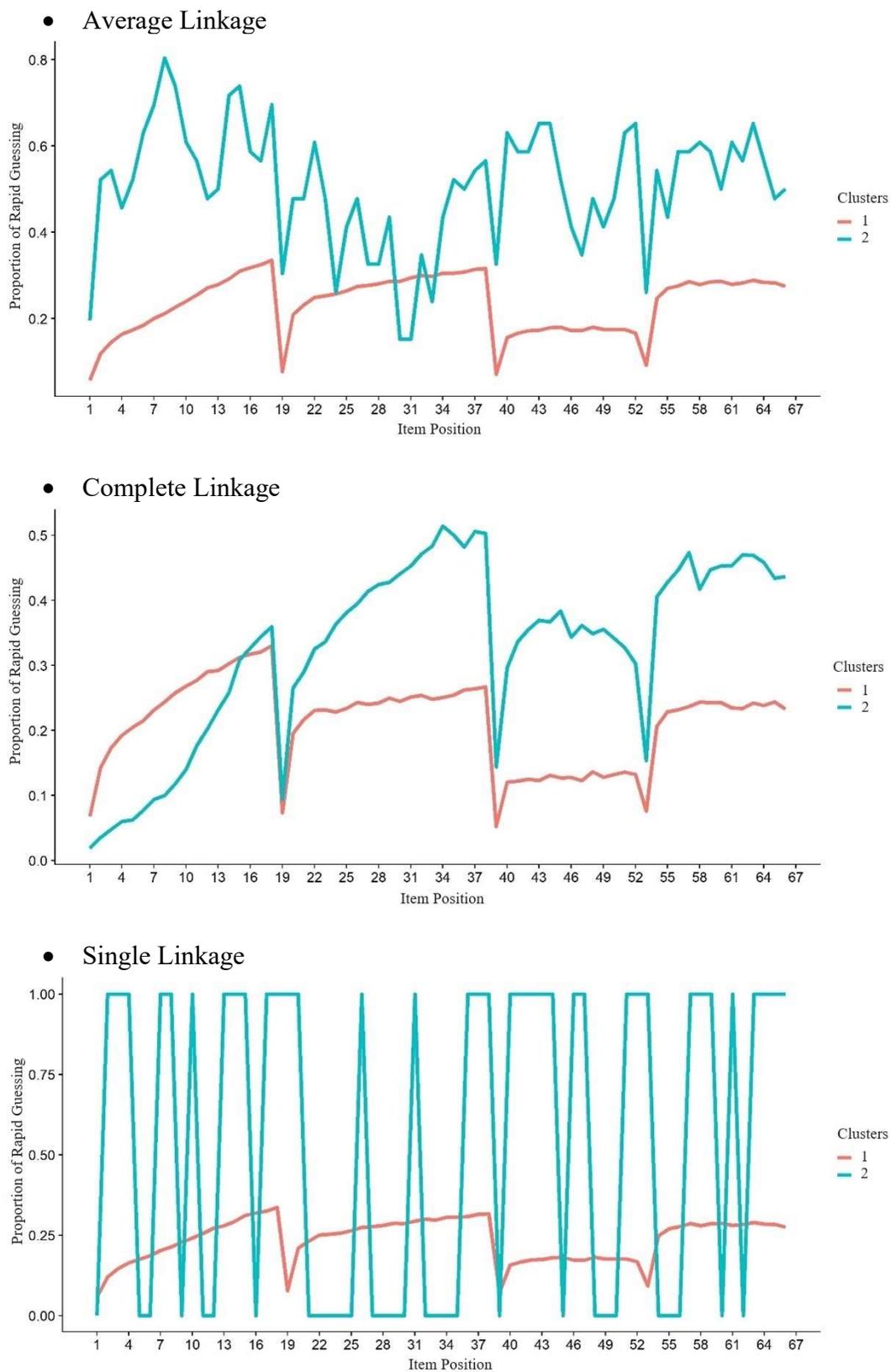

**Figure 5.** *Patterns from Aggregated Data versus Patterns from Data Separated by Rapid-Guessing Level (One Grade)*

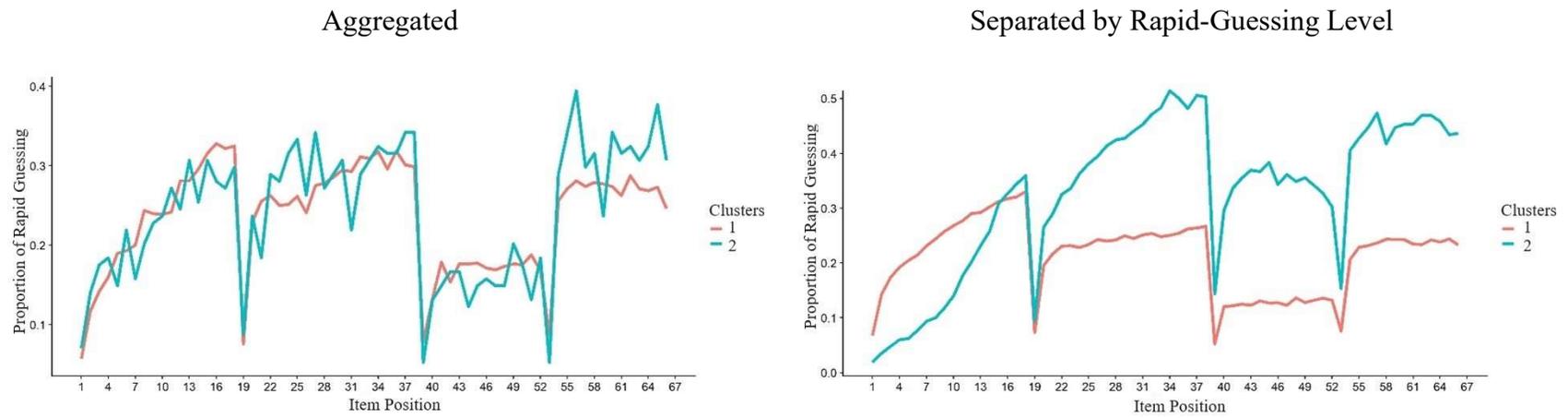

rapid-guessing patterns between the two clusters. We preferred to display rapid-guessing patterns for examinees of different rapid-guessing levels as they were simpler and more interpretable.

## Results

The results from the two random halves of the data (i.e., Set 1 and Set 2) corroborated those from all the data. In the following sections, we only report the results based on all for brevity. The optimal number of clusters was determined to be two for examinees in each rapid-guessing level per grade, based on the criteria of parsimony, consistency, and maximized between-cluster difference. A few consistent and highly interpretable patterns of rapid-guessing trajectories emerged from our analysis.

### Descriptive Statistics

Table 4 summarizes the descriptive statistics of the sample size and average number of rapid-guessing items for each grade, cluster, and rapid-guessing level. For the average number of rapid-guessing items, this table breaks down the information for the whole test ("All") and for each content domain separately ("D1" up to "D4").

The sample size varied widely across grades, clusters, and rapid-guessing levels, ranging from 1058 to 19811 for Cluster 1 and from 1255 to 17374 for Cluster 2. The average number of rapid-guessing items did not show any clear grade-level trends, but there were some interesting observations. First, as expected, examinees categorized as "high rapid guessing" had more rapid-guessing items ($\mu = 25.34$, $\sigma = 1.66$) than examinees classified as "moderate rapid guessing" ($\mu = 9.82$, $\sigma = 1.69$) across all grades and clusters. Second, the number of rapid-guessing items differed by domain, which suggested that some domains may be more prone to rapid guessing due to the unique features of this domain or its position in the test. Across all grades, clusters,

**Table 4.** *Descriptive Statistics: Sample Size and the Average Number of Item Positions Flagged with Rapid Guessing*

| Grade | Rapid-Guessing Level | Cluster 1 | | | | | | Cluster 2 | | | | | |
|---|---|---|---|---|---|---|---|---|---|---|---|---|---|
| | | $N$ | $\mu$ | | | | | $N$ | $\mu$ | | | | |
| | | | All | D1 | D2 | D3 | D4 | | All | D1 | D2 | D3 | D4 |
| 1 | Moderate | 19811 | 5.39 | 3.01 | 1.40 | 0.33 | 0.66 | 9120 | 10.54 | 1.30 | 4.29 | 1.96 | 2.99 |
| | High | 3658 | 25.63 | 10.26 | 8.74 | 3.00 | 3.63 | 4076 | 24.65 | 3.31 | 10.63 | 5.03 | 5.67 |
| 2 | Moderate | 5357 | 9.83 | 6.03 | 2.37 | 0.57 | 0.86 | 17374 | 7.66 | 1.90 | 3.10 | 1.09 | 1.58 |
| | High | 3425 | 24.28 | 9.49 | 8.96 | 2.69 | 3.13 | 3592 | 26.16 | 3.74 | 10.78 | 5.59 | 6.05 |
| 3 | Moderate | 7172 | 9.72 | 4.09 | 3.88 | 0.78 | 0.97 | 4887 | 7.62 | 1.10 | 2.54 | 1.52 | 2.46 |
| | High | 2189 | 24.04 | 8.59 | 9.34 | 2.82 | 3.30 | 1598 | 25.53 | 3.09 | 10.51 | 5.67 | 6.26 |
| 4 | Moderate | 8806 | 9.50 | 3.35 | 3.92 | 0.88 | 1.35 | 1577 | 10.81 | 1.06 | 2.69 | 2.51 | 4.55 |
| | High | 2122 | 24.35 | 8.68 | 9.8 | 2.68 | 3.19 | 1546 | 24.63 | 3.33 | 9.99 | 4.95 | 6.36 |
| 5 | Moderate | 1058 | 10.71 | 6.56 | 2.42 | 0.67 | 1.05 | 7742 | 10.16 | 2.36 | 3.81 | 1.46 | 2.53 |
| | High | 2215 | 24.20 | 7.95 | 9.00 | 3.04 | 4.21 | 1255 | 24.98 | 3.69 | 10.85 | 4.64 | 5.81 |
| 6 | Moderate | 8806 | 9.50 | 3.35 | 3.92 | 0.88 | 1.35 | 1577 | 10.81 | 1.06 | 2.69 | 2.51 | 4.55 |
| | High | 2122 | 24.35 | 8.68 | 9.80 | 2.68 | 3.19 | 1546 | 24.63 | 3.33 | 9.99 | 4.95 | 6.36 |
| 7 | Moderate | 6425 | 10.43 | 3.37 | 3.75 | 1.19 | 2.11 | 1638 | 11.66 | 1.23 | 2.19 | 2.65 | 5.58 |
| | High | 2821 | 24.25 | 8.12 | 9.20 | 2.75 | 4.17 | 1729 | 29.69 | 5.71 | 10.35 | 6.14 | 7.50 |
| 8 | Moderate | 1577 | 12.28 | 7.32 | 2.77 | 0.78 | 1.40 | 6992 | 10.51 | 1.95 | 3.44 | 1.73 | 3.40 |
| | High | 2083 | 28.92 | 11.79 | 9.70 | 3.09 | 4.35 | 3495 | 25.09 | 4.74 | 9.25 | 4.70 | 6.40 |

*Note.* *N* denotes sample size, and $\mu$ denotes the average of item positions flagged with rapid guessing.

and rapid-guessing levels, Domain 2 showed the highest number of rapid-guessing items ($\mu =$ 6.44, $\sigma = 3.50$), accounting for 32% of the 20 items in this domain. Domain 3 had the lowest number ($\mu = 2.69$, $\sigma = 1.71$), accounting for 19% of the 14 items in this domain. Third, the average number of rapid-guessing items varied in different ways for the two clusters, given the rapid-guessing level of the examinees. For examinees of "moderate rapid guessing," the average number of rapid-guessing items was higher for Cluster 2 than for Cluster 1 in half of the grades but not in the other half, showing no consistent pattern of difference between Cluster 1 and Cluster 2. For examinees of "high rapid guessing," on the contrary, Cluster 2 had a higher number of rapid-guessing items than Cluster 1 in all grades except Grade 1 and Grade 8.

**Rapid Guessing Patterns**

Figure 6 shows how the proportion of rapid-guessing items changed across item positions for different grades and rapid-guessing levels. Given the shared features, we started our pattern description by labeling the two clusters of examinees based on their rapid-guessing behavior to facilitate meaningful interpretation. Cluster 1, or "Frontloaded Guessers," had more rapid-guessing items in the first two domains, indicating low engagement at the start of the test. Cluster 2, or "Low-Start Climbers," had fewer rapid-guessing items in the first domain but more in the rest of the test, indicating high effort at the beginning but low effort later.

We identified some prominent patterns in rapid guessing. They are shared across grades or across rapid-guessing levels. Below are the details—

**Between-domain patterns throughout the test.** We observed that the trajectories reached a low point at the first item in each content domain, showing that only around 10% or fewer examinees had low effort in that item position. This suggests that examinees may pay more attention to the first item in a new domain, regardless of their cluster or grade. This could imply that the test design influenced the examinees' rapid guessing behavior. For example, the break that happened between domains could reset examinees' engagement.

The rapid-guessing rate was the lowest in Domain 3. This could indicate that all the examinees were somehow more engaged in Domain 3, possibly because the items in Domain 3 were more engaging or demanding than the items in other domains. Moreover, the rapid-guessing rate immediately jumped in Domain 4 after Domain 3. These patterns show that the examinees overall were not equally disengaged throughout the test, but rather they had some variation in their effort depending on the domain or its position in the test.

**Between-cluster variations in the first-half patterns.** There were variations between the two clusters of examinees in the patterns for the first half of the test. In the first half of the test, Frontloaded Guessers had a higher rapid-guessing rate than Low-Start Climbers in Domain 1 but not all the time for Domain 2; this rate increased steadily in Domain 1 for both clusters and in Domain 2 for Low-Start Climbers. This implies that both clusters became increasingly disengaged in Domain 1; Low-Start Climbers continued this trend in Domain 2 whereas Frontloaded Guessers did not. The rapid-guessing rate decreased in Domain 2 for Frontloaded Guessers, but increased for Low-Start Climbers, when compared with that in Domain 1. These patterns reveal that the two clusters reacted differently to Domain 2, while they reacted similarly to Domain 1. This could reflect their different attitudes or reaction mechanisms towards Domain 2 or its position in the test.

**Between-cluster variations in the second-half patterns.** The patterns for the second half of the test also differed for the two clusters of examinees. In the second half of the test, the rapid-guessing rate remained stable within each domain, indicating consistent disengagement for both clusters. However, Frontloaded Guessers had a lower rapid-guessing rate than Low-Start

Climbers in the second half of the test, reversing the pattern to some extent from the first half of the test. Similar to what we described in the prior interpretations of the first-half pattern variations, these second-half pattern variations could reflect their different attitudes or reaction mechanisms towards Domains 3 and 4 or their positions in the test.

**Grade bands associated with between-cluster variations in Domain 2.** We found it might be useful to provide more nuanced analysis and hence divided Grades 1-8 into three grade bands based on the differences in Domain 2. The three bands are shown in Figure 6, marked with three different background colors. The first grade band covers Grades 1-2 and 5. This band had a lower rapid-guessing rate for Frontloaded Guessers than for Low-Start Climbers in Domain 2. The second grade band covers Grades 3-4 and 6-7. It had the opposite pattern. The third grade band covers Grade 8, which had mixed patterns for the two rapid-guessing levels.

The subtle differences that resulted in grade bands reflect that examinees' rapid-guessing rate in Domain 2 varied by rapid-guessing level. For example, in the first grade band, Frontloaded Guessers were more likely to try or answer the items in Domain 2 than Low-Start Climbers. In the second grade band, Frontloaded Guessers were more likely to give up or skip the items in Domain 2 than Low-Start Climbers.

**Figure 6.** *Rapid-Guessing Trajectories in Three Grade Bands*

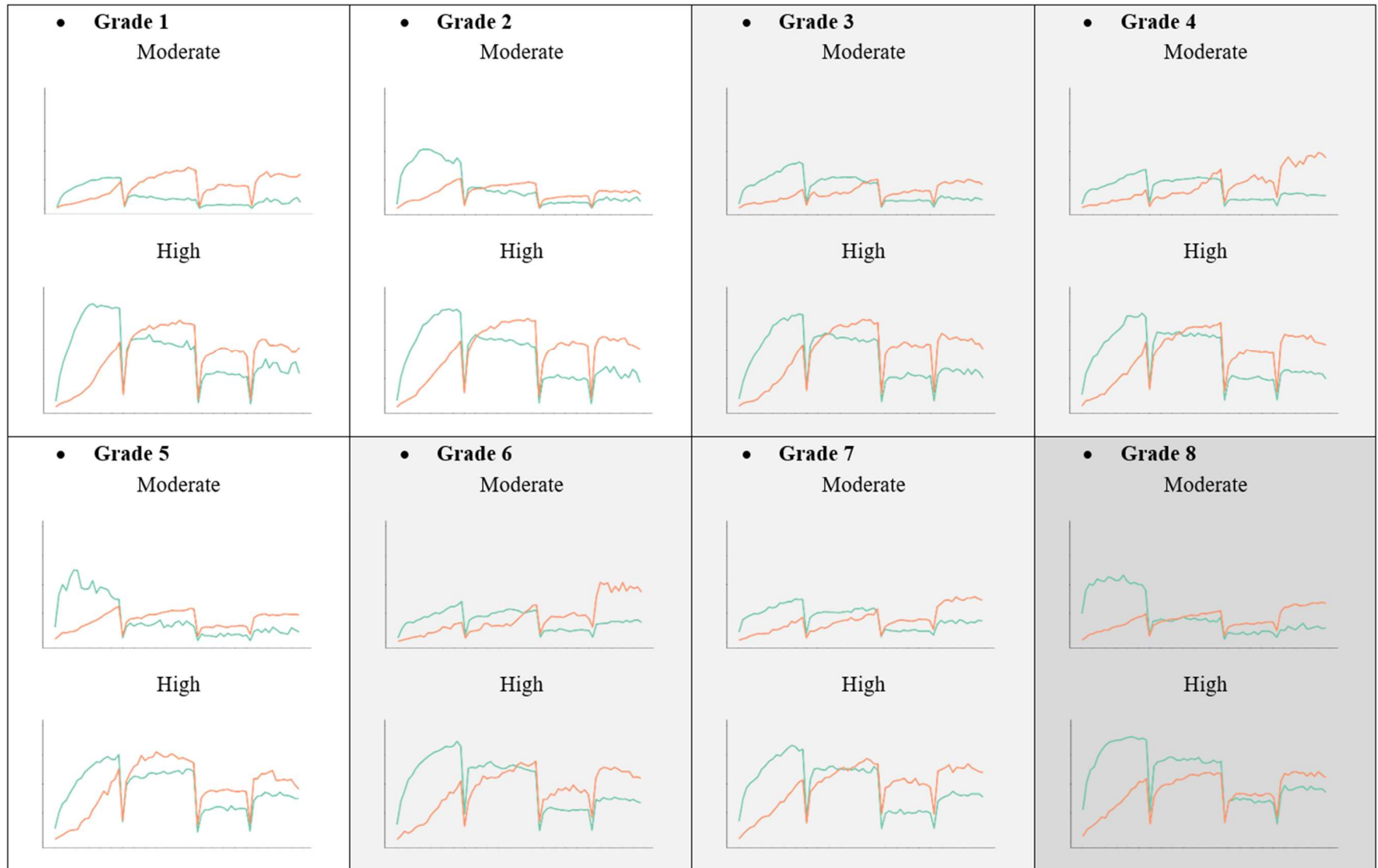

*Note.* There are three grade bands marked by three different shades.

## Discussion

In this study, we explored examinees' rapid-guessing response patterns in a CAT. Overall, our analysis unveiled interesting patterns that were not apparent or discussed in depth in the existing literature. These patterns suggest that there are underlying factors associated with rapid guessing in interim assessment and they can offer insights for test design and score validity for CAT. These factors include but are not limited to chronological grade, item position, content domain, "breaks," and test effort level. We recommend that test developers and administrators utilize hierarchical clustering to identify rapid-guessing patterns in CATs. By leveraging this data science technique, they can understand examinees' test-taking behaviors more effectively. Additionally, providing appropriate feedback, pacing, sequencing, incentives, and interventions based on these insights can help enhance examinee engagement in interim assessment.

We observed interesting trends in examinees' rapid guessing across different domains of the test. While we expected consistent effort across item positions, we found this pattern only in the latter half of the test, suggesting potential shifts in engagement levels. Also, we observed variations in engagement levels between different domains, with some domains eliciting higher engagement than others. In addition, all the trajectories had precipitous dips at adjoining points between domains. We speculate that "breaks" and/or content domain changes may have influenced examinees' engagement, either by resetting it or by requiring more cognitive adjustment (e.g., understanding level, testing strategies). We could not isolate the effect of each factor, because the dips coincided with "breaks" and content domain changes. However, we inferred that it may be "breaks" that have triggered examinees' re-engagement, especially for Domains 1 and 2, which were similar in content.

Moreover, there were cluster-associated variations in the rapid-guessing patterns throughout the test. For the two clusters of examinees we identified via hierarchical clustering, Frontloaded Guessers maintained and increased engagement throughout the test, whereas Low-Start Climbers started the test with high engagement but lost engagement as the test progressed. The nuanced patterns we described indicated that these two clusters might have had reaction mechanisms at different parts of the test, or may have possessed different attitudes, skills, or confidence towards the same test content. We speculate that Frontloaded Guessers may have gained more attention for completing Domain 2 after spending some time in the test, whereas Low-Start Climbers may have gotten even more disengaged after the starting phase.

Our study has implications for future research in related disciplines such as cognitive psychology. A possible direction is to examine the reasons behind the rapid guessing pattern differences for the two clusters of examinees. For example, future studies could explore why different clusters of examinees varied in their rapid guessing at different parts of a CAT. One possible way to do it is by comparing examinees' process data, such as cognitive reactions, when they progress in a CAT. Another approach is to combine hierarchical clustering results with qualitative insights from surveys or interviews with examinees, to help explain rapid-guessing patterns and factors that influence examinees' engagement. Such research would provide more insights into the test inclusiveness and fairness. Previous research on response time modeling, as reviewed in the Literature Review section, has not investigated the factors related to individual examinees' cognitive mechanisms. Future research could also explore how to use content or reaction constraints for test assembly or item selection. For example, the item selection algorithm could use adaptive constraints to account for both the examinee's ability and engagement when taking the test.

Our analysis has some limitations. We acknowledge that we did not account for the varied item characteristics that the adaptive format of the CAT may present at identical item positions, a nuance that can further complicate the rapid-guessing patterns in our study. However, we aggregated the rapid-guessing information at the same item positions, regardless of item characteristics, to elicit position-centric patterns that are relevant to test developers and valid for our interpretations. Also, we derived the patterns for each chronological grade, assuming that they would have practical implications for test design and instruction, since grades are the common units for testing and teaching. However, examinees in the same grade may have different competencies and reporting results by examinees' ability level may offer another perspective. That said, the ability estimates may also be inaccurate if the rapid-guessing responses are not taken care of in the first place. Moreover, our results may not apply to the entire CAT population, as they were based on the data from a specific group of examinees who showed rapid-guessing responses when taking the test in school and in one sitting during Fall 2021, not representing all the examinees.

Our interpretation of results based on clustering the data led to many insights but has inherent limitations. For example, our hierarchical clustering analysis reveals associations between variables, which cannot establish causation or definitively explain underlying motivations. However, our observations can inspire further investigation, to unveil whether examinees may truly exhibit low motivation or different reaction mechanisms associated with test design factors, as we speculated. To address this limitation, future research could employ complementary qualitative methods such as surveys, interviews, or observational studies to provide richer context and validate our results. By integrating qualitative data, we can better understand the complexities of examinees' rapid-guessing behavior and enhance the robustness of our findings.

## Conclusion

According to Alber (2017), a classroom teacher, "low-stakes assessments are really the most important and useful student data." Our study was intended to address the existing literature gap by investigating rapid-guessing patterns in CAT-driven interim assessment. Via hierarchical clustering, our findings unveiled that "breaks" or the change of the content domain was associated with examinees' effort, and different types of examinees showed different rapid-guessing patterns. While our findings were based on correlations, which do not establish causations, they provide valuable insights into the relationship between variables and warrant further investigation through experimental or quasi-experimental research designs. Also, we demonstrated a novel approach by employing hierarchical clustering to analyze process data (i.e., item response time) in testing.